\documentclass[11pt]{article}

\usepackage[preprint]{acl}
\usepackage{times}
\usepackage{latexsym}
\usepackage{scrextend}
\usepackage[T1]{fontenc}

\usepackage[utf8]{inputenc}
\usepackage{booktabs}
\usepackage{microtype}

\usepackage{inconsolata}

\usepackage{graphicx}
\usepackage{amsmath}
\usepackage{multirow}
\usepackage{tcolorbox}
\tcbuselibrary{listings, breakable}

\definecolor{promptbackground}{RGB}{235, 245, 255}
\definecolor{promptframe}{RGB}{60, 120, 180}
\definecolor{outputbackground}{gray}{0.95}
\definecolor{outputframe}{gray}{0.65}
\newtcblisting{promptbox}[2][]{
    listing only, breakable, title=#2,
    colback=promptbackground, colframe=promptframe, coltitle=white,
    fonttitle=\bfseries, boxrule=0.5pt, arc=3mm, colbacktitle=promptframe,
    fontupper=\small\ttfamily,
    #1
}

\title{Do Coding Agents Reuse Existing Code or Reinvent the Wheel?}

\author{
    \textbf{Dongsheng Ma}$^\dagger$\textsuperscript{1},
    \textbf{Sizhe Wang}$^\dagger$\textsuperscript{2},
    \textbf{Xinyi Huang}\textsuperscript{1},
    \textbf{Zhengren Wang}\textsuperscript{\textbf{3}} \\
    \textbf{Yuhan Wang}\textsuperscript{\textbf{1}},
    \textbf{Luyang Si}\textsuperscript{\textbf{4}},
    \textbf{Xincheng Wei}\textsuperscript{\textbf{5}},
    \textbf{Wentao Zhang}\textsuperscript{*\textbf{1,6}}\\
    \textsuperscript{1}Peking University
    \textsuperscript{2}Fudan University
    \textsuperscript{3}Shanghai Jiao Tong University
    \textsuperscript{4}Tsinghua University\\
    \textsuperscript{5}The Chinese University of Hong Kong, Shenzhen
    \textsuperscript{6}Zhongguancun Academy\\
    \texttt{madongsheng26@stu.pku.edu.cn}, 
    \texttt{wangsz26@m.fudan.edu.cn}, 
    \texttt{wentao.zhang@pku.edu.cn}\\
}

\begin{document}
\maketitle
\deffootnote[1.5em]{1.5em}{1em}{}
\renewcommand{\thefootnote}{\fnsymbol{footnote}}
\footnotetext{$\dagger$ Equal contribution (listed alphabetically);\\
* Corresponding author. Work In Progress.
}
\renewcommand{\thefootnote}{\arabic{footnote}}

\begin{abstract}
Coding agents are increasingly deployed for iterative development on real repositories, yet existing evaluation barely answers a basic question: \emph{do coding agents reuse existing code or reinvent the wheel?} The question matters: every duplicated implementation is a fix applied twice and agents produce code far faster than humans can audit, so redundancy accumulates unsupervised. Thus, we present \textbf{RepoReuse}, a multi-turn benchmark for auditing code reuse in real repositories, where requirements are revealed turn by turn and the workspace accumulates across turns. It is built by a fully automated pipeline combining AST-based dependency graphs, guided evidence collection, and execution-verified task synthesis, and scales readily to new repositories. Beyond pass rates, we measure the reuse rate together with recall and cross-turn structural redundancy. An audit over 3{,}000 turns shows that agents progressively stop exploring relevant repository code, reuse their own history less even when it is fully in the workspace, and leave duplicated logic in 50.8\% of task chains by turn~5---all while pass rates barely move. Such deficiencies are invisible to pass rates, underscoring the need to evaluate code generation beyond functional correctness.
\end{abstract}

\section{Introduction}
\label{sec:intro}

Coding agents continue to break records on coding benchmarks~\citep{kimiteam2026kimik3openfrontier,deepseekai2026deepseekv41flashpushinglimitskv,yang2024sweagentagentcomputerinterfacesenable,wang2025openhandsopenplatformai,jimenez2024swebenchlanguagemodelsresolve}, and are increasingly deployed in real repositories to assist humans with iterative development. Yet passing tests is only half of what makes code good. Code is read, extended, and maintained long after it is written; a solution that \emph{reinvents the wheel} is not a good solution, even when every test is green.

\textbf{Motivation} Redundancy is not a cosmetic issue. Over time, every duplicated implementation means a bug fix or an interface change that must be applied twice, a piece of functionality that silently drifts between parallel copies, and more code that every later turn---human or agent---must wade through. Recent studies have already observed the signs: in long-horizon iterative tasks, the redundancy of agent-produced code keeps rising, and agent-generated patches are markedly longer than reference solutions that reuse existing modules~\citep{Orlanski2026SlopCodeBench,li2026loopsbenchharnessengineeringloop,abbassi2025taxonomyinefficienciesllmgeneratedpython,liu2025codecopycatconundrumdemystifying}. What is truly alarming is the asymmetry of speed: agents write code far faster than human experts can audit it, so this failure mode accumulates at a staggering rate, quietly piling up into ``spaghetti code.''

\begin{figure*}[t]
    \centering    \includegraphics[width=1.0\linewidth]{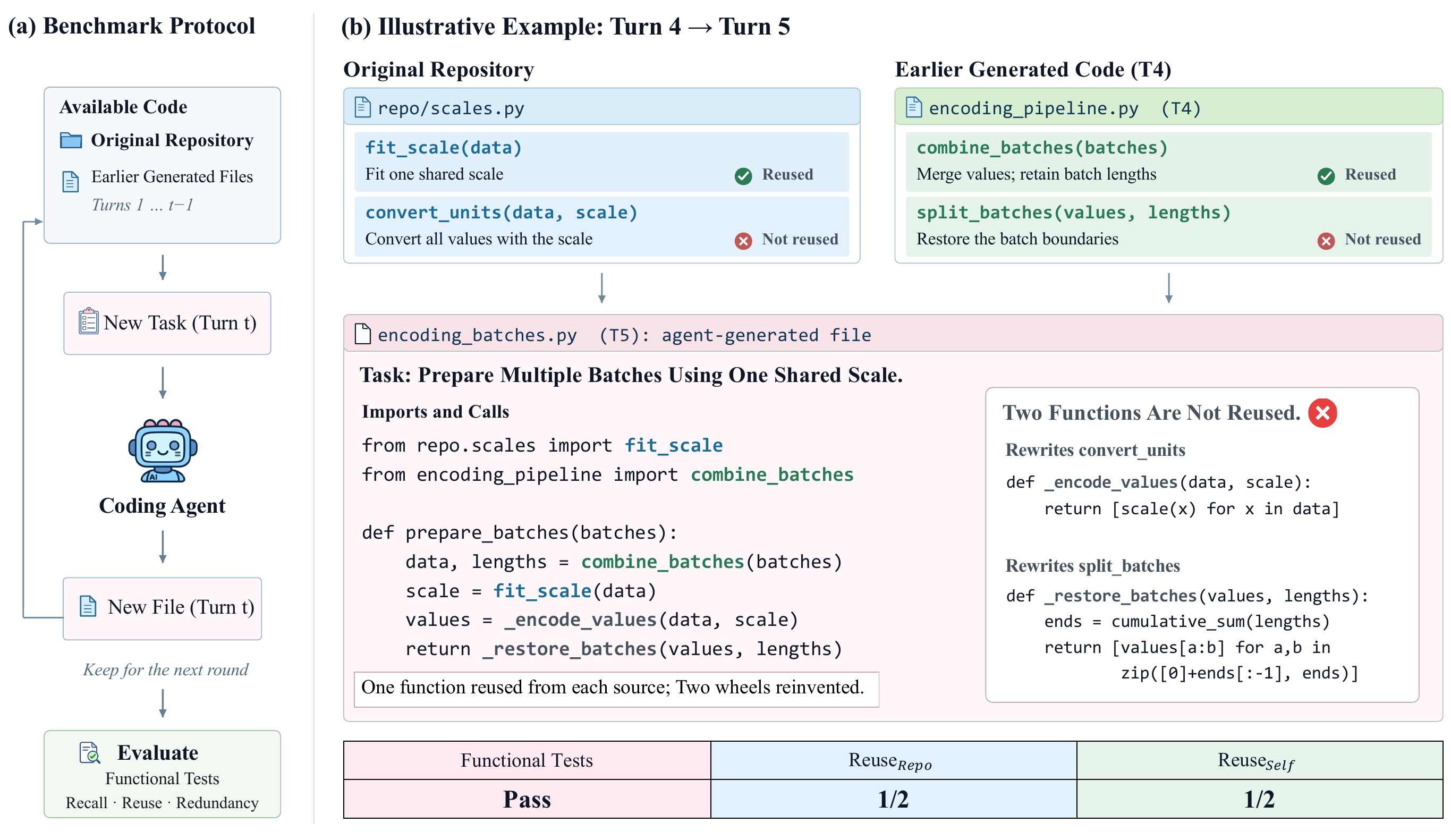}
    \caption{\textbf{Overview of RepoReuse.} \emph{Left:} the benchmark protocol---turn-wise requirements over a workspace that accumulates the agent's own code, scored per turn by functional tests and by recall, reuse, and redundancy. \emph{Right:} an illustrative example---the agent reuses one repository symbol and one of its own earlier functions but rewrites the two remaining targets; tests pass while both reuse rates are $1/2$.}
    \label{Overview}
\end{figure*}

Yet our evaluation instruments are, in principle, blind to all of this. Existing benchmarks are built almost entirely around functional correctness: HumanEval~\citep{chen2021evaluatinglargelanguagemodels}, MBPP~\citep{austin2021programsynthesislargelanguage}, and SWE-bench~\citep{jimenez2024swebenchlanguagemodelsresolve} score only whether tests pass (i.e., the pass rate), paying little attention to another key dimension of how models write code---\emph{reusability}. Prior work touches on this dimension but does not measure it in realistic multi-turn, repository-scale development: benchmarks on reuse and maintainability either iterate only at the function level~\citep{wang-etal-2026-codeflowbench} or are confined to a single turn~\citep{wang2025maintaincodermaintainablecodegeneration}, while repository-level efforts stop at one-shot completion or well-documented entry points rather than the discovery and reuse of internal modules~\citep{ding2023crosscodeevaldiversemultilingualbenchmark,liu2023repobenchbenchmarkingrepositorylevelcode,tang2024mlbenchevaluatinglargelanguage}.

Real iterative development unfolds precisely at the repository scale, across turns: a model must keep developing on an existing multi-file repository, \emph{reusing existing code}---not only the repository's own modules, but also the code it wrote in previous turns, which has by then become part of the repository. To this end, we construct \textbf{RepoReuse}, a multi-turn benchmark for auditing code reuse in real repositories (See Figure~\ref{Overview}). Requirements are revealed turn by turn and the workspace accumulates across turns, so that an agent's reuse of both pre-existing repository modules and its own historical implementations can be measured turn by turn. RepoReuse is built by a fully automated pipeline that combines AST-based dependency graphs, guided evidence collection, and execution-verified task synthesis, and scales readily to new repositories~\citep{le2026sweevobenchmarkingcodingagents,jain2024livecodebenchholisticcontaminationfree}.

Our metrics mirror the agent's per-turn workflow: reading code, deciding what to reuse, and writing new code. The central metric, \emph{reuse rate}, asks whether the submission calls each target through a real call edge, identified by AST analysis against the reference solution's actual dependencies; repository and self-produced targets are measured separately ($\text{reuse}_{\text{repo}}$ / $\text{reuse}_{\text{self}}$). To explain why a target is missed, upstream \emph{recall} records the fraction of target source code read in the current turn, separating exploration-side gaps (never read) from execution-side gaps (read but not used); downstream, $C_{\text{dup}}$ counts targets that are not called yet have most of their logic rewritten, capturing the redundancy they leave behind~\citep{4222572}. Together, the three trace a reuse decision from exploration to consequence, forming a complete auditing instrument for code reuse.

With this auditing instrument, we uncover systematic reuse deficiencies of coding agents. On the exploration side, agents gradually stop retrieving as turns progress: they read 83.6\% of relevant repository code at turn~1 on average, but only 35.4\% at turn~5. Once the workspace is filled with their own code, reuse decisions are increasingly made from signatures and priors rather than grounded in current-turn observation. On the execution side, a more counterintuitive phenomenon is that seeing is not using: recall over their own historical code is nearly saturated in every turn, yet self-reuse still decays from 83.9\% to 69.1\%. Ablations further show that disclosing the complete source of historical implementations barely helps, performing on par with providing no memory at all---the bottleneck is not access to information, but the agent's own disposition to build on existing code. Downstream, bypassed targets steadily accumulate as structural redundancy in the workspace: the fraction of task chains containing a cross-turn re-implementation climbs from 13.8\% at turn~1 to 50.8\% at turn~5, while pass rates barely move. Agents solve task after task correctly while the codebase grows messier turn after turn, pass-rate-centered evaluation is structurally blind to all of it.

Our contributions are as follows:
\begin{itemize}
    \item \textbf{An important question:} we raise a question neglected by existing evaluation: \emph{do coding agents reuse existing code or reinvent the wheel?} We give the first systematic, turn-by-turn measurable answer: agents steadily accumulate redundant code that piles up into ``spaghetti code.''
    \item \textbf{The RepoReuse benchmark:} a repository-level multi-turn iterative development benchmark with turn-wise requirement revelation, measuring an agent's reuse of both pre-existing repository code and its own historical code turn by turn. It is built by a fully automated pipeline that requires no human involvement and scales readily to new repositories.
    \item \textbf{Auditing findings:} a systematic audit over 3{,}000 turns of development shows that reuse deficiencies are pervasive and deepen over turns: agents see their own historical code yet fail to reuse it (self reuse decays from 83.9\% to 69.1\% even under saturated recall); by turn~5, half of the task chains have accumulated cross-turn re-implementations while pass rates barely move---deficiencies entirely invisible to pass-rate-centered evaluation.
\end{itemize}

\section{Related Work}
\label{sec:related}

\paragraph{Coding Agents and Harnesses}Coding agents are moving from research prototypes into real development workflows: new-generation models (e.g., Kimi K3~\citep{kimiteam2026kimik3openfrontier}, DeepSeek v4.1~\citep{deepseekai2026deepseekv41flashpushinglimitskv}) treat agentic coding as a core capability, while harnesses range from research frameworks (SWE-agent~\citep{yang2024sweagentagentcomputerinterfacesenable}, OpenHands~\citep{wang2025openhandsopenplatformai}) to industrial products (Claude Code~\citep{anthropic2025claudecode}, Codex~\citep{codex}, DeepSeek Harness~\citep{deepseek-harness2026}).
It is increasingly clear that harness and interaction design have become performance variables on par with the backbone model~\citep{vats2026scaffoldeffectcodingagents}:
SWE-agent highlights the role of tool-interface design, and Agentless~\citep{xia2024agentlessdemystifyingllmbasedsoftware} shows that a simple localization--repair--validation pipeline is competitive even without an autonomous interaction loop.
Yet evaluation along this line is defined almost entirely by task success rate---\emph{how} an agent writes code (whether it reuses existing implementations or piles up redundancy) is not measured. RepoReuse instead audits the reuse behavior of existing agents and harnesses.

\paragraph{Code Generation Evaluation}Code generation evaluation has long centered on functional correctness, from function-level tests (HumanEval~\citep{chen2021evaluatinglargelanguagemodels}, MBPP~\citep{austin2021programsynthesislargelanguage}) to issue resolution in real repositories (SWE-bench~\citep{jimenez2024swebenchlanguagemodelsresolve}) and repository-level completion (CrossCodeEval~\citep{ding2023crosscodeevaldiversemultilingualbenchmark}, RepoBench~\citep{liu2023repobenchbenchmarkingrepositorylevelcode}, ML-Bench~\citep{tang2024mlbenchevaluatinglargelanguage}).
Closest to us are multi-turn benchmarks: CodeFlowBench~\citep{wang-etal-2026-codeflowbench} iterates within a single file; MaintainCoder~\citep{wang2025maintaincodermaintainablecodegeneration} varies requirements but remains single-turn; SWE-Chain~\citep{lam2026swechainbenchmarkingcodingagents} asks agents to upgrade a package step by step along its version chain, each step building on the agent's own previous codebase; EvoCode-Bench~\citep{shen2026evocodebenchevaluatingcodingagents} and SWE-EVO~\citep{le2026sweevobenchmarkingcodingagents} evaluate multi-turn iterative interaction and long-horizon version evolution under per-turn functional tests; LoopsBench~\citep{li2026loopsbenchharnessengineeringloop} organizes each task as a dependency DAG with tests released along the ready frontier, observing bloated patches and prerequisite dependencies missed by agent plans; SlopCodeBench~\citep{Orlanski2026SlopCodeBench} lets agents repeatedly extend their own solutions under evolving specifications, observing structural erosion and rising redundancy; SWE-Explore~\citep{zhang2026sweexplorebenchmarkingcodingagents} measures line-level exploration recall, but over static snapshots with execution deliberately removed.
These works measure outcomes and artifacts, and none audits models' reuse behavior in realistic repository-level multi-turn development. RepoReuse fills this gap: under a repository-level multi-turn iterative setting, it measures, turn by turn, an agent's reuse of both pre-existing repository code and its own historical code.

\section{RepoReuse}
\label{sec:reporeuse}

\subsection{Benchmark Construction}

Constructing RepoReuse requires delivering two things at once: multi-turn task chains, and a ground-truth answer to ``what should be reused at each turn.'' To this end, we implement a fully automatic construction pipeline: starting from mature open-source repositories, it produces tasks through three stages---evidence collection, task synthesis, and execution verification---without any human involvement. The pipeline guarantees that the reference solution genuinely reuses the designated modules and that the test cases come from actual execution, so that reuse auditing rests on verifiable ground truth.

\paragraph{Repository Sources}
RepoReuse builds its tasks on mature, widely used open-source Python libraries spanning domains such as scientific computing and data visualization. Selected repositories must satisfy three conditions: (1) a multi-file pure-Python codebase with a rich internal implementation layer beneath the public API---the evidence for our tasks comes from these undocumented internal subpackages rather than well-documented public interfaces; (2) a reproducible environment: each repository is pinned to a specific commit, and its dependency installation recipe is directly inherited from SWE-bench's~\citep{jimenez2024swebenchlanguagemodelsresolve} install table, ensuring consistency between the construction and evaluation environments.

\paragraph{Dependency Graph and Evidence Collection}
We first build a callable ``map'' of the repository: via AST-based static analysis, functions and methods are represented as nodes carrying precise line ranges, signatures, and docstrings, with call relations as edges, forming a dependency tree that covers the whole repository; internal symbols beneath the public API surface are marked as candidate starting points. A strong model (GPT-5.6 Sol~\citep{openai_gpt56}) then performs guided walks over the dependency tree: starting from an internal symbol, it selects the next symbol hop by hop with justifications, collecting a group of interdependent modules as an \emph{evidence pack}~\citep{li2025websailornavigatingsuperhumanreasoning,wu2025webdancerautonomousinformationseeking,ma2026citevqabenchmarkingevidenceattribution,zhang2026docdanceragenticdocumentgroundedinformation}. Walks from turn 2 onward depart from the previous turn's artifact, making the reuse of earlier-turn functions a natural requirement of the task chain; to prevent evidence packs from degenerating into self-wrapping over turns, an additional nearby but previously unused repository symbol is injected during each walk.

\paragraph{Task Synthesis}
Based on the evidence pack, the authoring model (GPT-5.6 Sol) synthesizes the complete content of a turn: (1) a \emph{requirement} that describes only the target behavior and never names any evidence-pack symbol---otherwise exploration degenerates into a string search; (2) a \emph{reference solution} that must genuinely call the evidence-pack modules and the chain's previous-turn artifacts; (3) \emph{test cases} whose expected values are taken from actual runs of the reference solution rather than the authoring model's claimed outputs. Modules that pass verification are written back into the repository package under semantic names, side by side with real code, for later turns to depend on.

\paragraph{Verification}
The core of verification is the reliability of the test cases themselves: every expected value is obtained by actually executing the reference solution, and cases that raise errors or produce unstable outputs are discarded; if too many are discarded for the remaining cases to cover the declared functions, the whole turn is regenerated. On top of this, each turn must pass automatic checks---e.g., the reference solution passes all tests, an empty implementation must fail, and AST analysis confirms that the reference solution genuinely reuses the evidence-pack modules. Finally, an independent audit, trusting none of the generator's outputs, reinstalls the entire task chain and re-runs all tests before the task is packaged for delivery.

\subsection{Benchmark Overview}

Table~\ref{tab:stats} summarizes the scale and composition of RepoReuse. Each turn ships with 10.8 execution-verified test cases on average; from turn 2 onward, the reference solution reuses 1.9 pre-existing repository symbols and 2.4 functions produced in earlier turns per turn---both types of reuse targets genuinely exist in every turn, providing a stable measurement basis for reuse auditing.

\begin{table}[t]
\centering
\caption{Statistics of RepoReuse.}
\label{tab:stats}
\begin{tabular}{@{}lc@{}}
\toprule
Statistic & Value \\
\midrule
Tasks & 75 \\
Avg.\ turns per task & 5.0 \\
Avg.\ test cases per turn & 10.8 \\
Avg.\ requirement length (chars) & $\sim$2{,}570 \\
New functions per turn & 3 \\
Reused symbols per turn (repo / self) & 1.9 / 2.4 \\
\bottomrule
\end{tabular}
\end{table}

\subsection{Evaluation Framework}
\label{sec:framework}

In each turn, a coding agent reads existing code, decides which modules to reuse, and generates new code. We first formalize this multi-turn setting and the reuse targets it induces, and then define metrics organized around reuse along the same workflow.

\subsubsection{Task Formulation}
Let $W_0$ denote the initial repository. At each turn $t$, the agent receives a requirement $q_t$ and the current workspace $W_{t-1}$, and produces a submission $s_t$:
\begin{equation*}
    (q_t,\; W_{t-1}) \;\xrightarrow{\text{agent}}\; s_t, \qquad W_t = W_{t-1} \oplus s_t
\end{equation*}
where $\oplus$ denotes installing the submission into the workspace. The agent's own submissions, not the reference solutions, are carried into later turns, so every turn builds on what the agent itself wrote. For each turn, we extract the reuse targets from the actual dependencies of the reference solution via AST analysis and split them into two disjoint sets: \emph{repo targets} $\mathcal{T}^{\text{repo}}_t \subseteq W_0$, functions present in the original repository, and \emph{self targets} $\mathcal{T}^{\text{self}}_t \subseteq \{s_1, \dots, s_{t-1}\}$, functions produced by the agent in earlier turns. Self targets exist only from turn~2 onward.

\subsubsection{Metric Definition}
Our metrics mirror the per-turn workflow. The central \emph{reuse} metric asks whether the agent adopts each target. \emph{Upstream} metrics describe what the agent explored before writing code, and help explain why a target was or was not reused. \emph{Downstream} metrics describe the outcome, both whether the new code works and what bypassing a target leaves in the workspace.

\paragraph{Reuse Metric}
For $s \in \{\text{repo}, \text{self}\}$, the reuse rate is the fraction of targets that the submission directly calls:
\begin{equation}
    \text{reuse}_{s}(t) = \frac{|\{f \in \mathcal{T}^{s}_t : f \text{ is called in } s_t\}|}{|\mathcal{T}^{s}_t|}
\end{equation}
A call is identified by a real call edge to the target through AST-based value-reference name matching. Each target is scored 0/1 regardless of invocation count, and a function the agent redefines locally under the same name does not count as a call. For example, $\text{reuse}_{\text{self}}(t) = 0.5$ means that the submission calls half of the earlier-turn functions the reference solution builds on and bypasses the other half.

\paragraph{Upstream Metrics}
Reuse rate tells us \emph{what} happened but not \emph{why}: a target that was not reused may never have been found, or may have been found but not adopted. To separate the two, we measure \emph{recall}. A target is recalled if the agent's file-read actions in the current turn overlap at least one line of its source body; directory listings do not count as reads:
\begin{equation}
    \text{recall}_{s}(t) = \frac{|\{f \in \mathcal{T}^{s}_t : f \text{ is read in turn } t\}|}{|\mathcal{T}^{s}_t|}
\end{equation}
Combined with reuse, recall sorts every missed target into one of two cases. A target that was read but not reused points to an \emph{execution-side} gap, where the agent saw the code but chose not to use it; a target neither read nor reused points to an \emph{exploration-side} gap, where the agent never found it. To characterize how much the agent explores, we also report two behavior statistics per turn: \emph{Files}, the number of distinct files viewed, searched, or listed, and \emph{Lines}, the number of distinct source lines viewed. These statistics are descriptive rather than evaluative and help interpret recall differences across models and harnesses.

\paragraph{Downstream Metrics}
We first retain standard functional correctness. \emph{Pass rate} is the proportion of test cases passed in a turn, and \emph{resolve rate} is the proportion of turns in which all tests pass. Correctness, however, captures only one consequence of reuse. As agents produce code faster than humans can audit it, an agent may solve each task correctly while repeatedly re-implementing functionality that already exists, quietly turning the codebase into redundant ``spaghetti code.'' We therefore also measure whether the agent \emph{re-implements} existing code~\citep{4222572}. A target $f \in \mathcal{T}^{\text{repo}}_t \cup \mathcal{T}^{\text{self}}_t$ is re-implemented at turn $t$ if $s_t$ does not call $f$ but calls at least 80\% of the non-builtin functions that $f$ itself calls, considering only targets that call at least three such functions. Because every turn inherits the workspace of earlier turns, a single re-implementation leaves a redundant copy for the rest of the task chain; we call any chain containing one a \emph{duplicated chain}, and measure the \emph{duplicated-chain rate}, the fraction of task chains $\mathcal{C}$ that have become duplicated by turn $t$:
\begin{equation}
    C_{\text{dup}}(t) = \frac{1}{|\mathcal{C}|} \sum_{c \in \mathcal{C}} \mathbf{1}\big[\exists\, i \le t:\ \mathcal{R}_{c,i} \neq \emptyset\big]
\end{equation}
where $\mathcal{R}_{c,i}$ is the set of re-implemented targets at turn $i$ of chain $c$. $C_{\text{dup}}$ is cumulative, never decreases along a chain, and is better when lower.

\section{Experiments}
\label{sec:exp}

\subsection{Experiment Setup}
\label{sec:setup}

\paragraph{Models and Harnesses}
Since both the backbone model and the harness shape how an agent explores and writes code~\citep{vats2026scaffoldeffectcodingagents}, we evaluate a grid of model--harness combinations. We use two representative harnesses: \textbf{mini-SWE-agent}~\citep{minisweagent2024}, a minimal harness whose only action is a bash command, and \textbf{OpenCode}~\citep{opencode2026}, an open-source industrial harness with native read, grep, glob, edit, and write tools.
We pair them with four backbone models: GPT-5.6 Terra~\citep{openai_gpt56}, DeepSeek-v4.1-flash~\citep{deepseekai2026deepseekv41flashpushinglimitskv}, Qwen3.7-plus~\citep{qwen2026qwen37plus}, and GLM-5.3~\citep{zai2026glm53}.

\paragraph{Evaluation Protocol}
All runs follow the self-invoking protocol: each turn starts a fresh agent session, while the workspace persists across turns and keeps the agent's own earlier implementations; a failed turn does not end the chain. From turn~2 onward, the requirement lists the public interfaces of the functions the agent implemented in earlier turns but not their source code, and repository reuse targets are never named. All configurations share the same per-turn budget. All metrics are macro-averaged over turns where they are defined. 


\subsection{Main Results}
\label{sec:main-results}

\begin{table*}[t]
\centering
\small
\setlength{\tabcolsep}{4pt}
\caption{Main results on RepoReuse. Columns are organized around reuse: upstream metrics describe how the agent explores, and downstream metrics describe the outcome. Reuse, recall, correctness, and $C_{\text{dup}}$ are in \%; $C_{\text{dup}}$ is averaged over the five cumulative turn-wise rates. Best reuse, recall, and correctness per column in \textbf{bold}.}
\label{tab:main}
\begin{tabular}{@{}llccccccccc@{}}
\toprule
& & & & \multicolumn{4}{c}{Upstream} & \multicolumn{3}{c}{Downstream} \\
\cmidrule(lr){5-8}\cmidrule(lr){9-11}
& & \multicolumn{2}{c}{Reuse} & \multicolumn{2}{c}{Recall} & \multicolumn{2}{c}{Agent Behavior} & \multicolumn{2}{c}{Correctness} & Redundancy \\
\cmidrule(lr){3-4}\cmidrule(lr){5-6}\cmidrule(lr){7-8}\cmidrule(lr){9-10}\cmidrule(lr){11-11}
Harness & Model & repo & self & repo & self & Files & Lines & Resolve & Pass & $C_{\text{dup}}$ \\
\midrule
\multirow{5}{*}{mini-SWE-agent}
& GPT-5.6 Terra     & 62.8 & 71.1 & 52.7 & 99.6 & 20.5 & 1150 & 48.8 & 81.2 & 42.7 \\
& DeepSeek-v4.1-flash & 68.8 & 84.2 & \textbf{61.4} & 99.9 & 45.4 & 1267 & 66.4 & 90.8 & 21.1 \\
& Qwen3.7-plus      & 48.9 & 67.7 & 36.2 & 99.6 & 22.0 & 680 & 27.7 & 71.1 & 37.1 \\
& GLM-5.3           & 55.3 & 71.7 & 47.2 & 99.2 & 36.5 & 1104 & 44.3 & 81.2 & 39.7 \\
\midrule
\multirow{5}{*}{OpenCode}
& GPT-5.6 Terra     & 61.6 & 67.0 & 42.3 & \textbf{100.0} & 18.5 & 479 & 44.5 & 80.1 & 36.8 \\
& DeepSeek-v4.1-flash & \textbf{75.7} & \textbf{86.4} & \textbf{61.4} & 98.9 & 44.1 & 1098 & \textbf{72.3} & \textbf{91.6} & 17.1 \\
& Qwen3.7-plus      & 50.5 & 67.8 & 35.1 & 99.7 & 13.5 & 576 & 30.4 & 71.9 & 46.7 \\
& GLM-5.3           & 56.1 & 73.7 & 51.5 & 99.9 & 29.3 & 900 & 44.3 & 81.0 & 33.6 \\

\bottomrule
\end{tabular}
\end{table*}

\begin{table*}[t]
\centering
\small
\setlength{\tabcolsep}{4pt}
\caption{Recall, reuse, and redundancy at the first and last turn (\%). Self reuse is defined from turn~2; $\Delta$ is the change from the first to the last turn. Self reuse declines while the share of task chains containing re-implemented targets grows over turns.}
\label{tab:turns}
\begin{tabular}{@{}llcccccccccccc@{}}
\toprule
& & \multicolumn{3}{c}{$\text{recall}_{\text{repo}}$} & \multicolumn{3}{c}{$\text{reuse}_{\text{repo}}$} & \multicolumn{3}{c}{$\text{reuse}_{\text{self}}$} & \multicolumn{3}{c}{$C_{\text{dup}}$} \\
\cmidrule(lr){3-5}\cmidrule(lr){6-8}\cmidrule(lr){9-11}\cmidrule(lr){12-14}
Harness & Model & T1 & T5 & $\Delta$ & T1 & T5 & $\Delta$ & T2 & T5 & $\Delta$ & T1 & T5 & $\Delta$ \\
\midrule
mini-SWE-agent & GPT-5.6 Terra     & 86.3 & 38.8 & $-$47.5 & 61.7 & 64.3 & +2.6 & 77.1 & 69.0 & $-$8.1 & 14.7 & 60.0 & +45.3 \\
mini-SWE-agent & DeepSeek-v4.1-flash & 88.1 & 49.8 & $-$38.3 & 74.4 & 72.2 & $-$2.2 & 92.2 & 79.3 & $-$12.9 & 10.7 & 33.3 & +22.6 \\
mini-SWE-agent & Qwen3.7-plus      & 80.1 & 21.7 & $-$58.4 & 52.5 & 50.1 & $-$2.4 & 84.2 & 61.3 & $-$22.9 & 14.7 & 57.3 & +42.6 \\
mini-SWE-agent & GLM-5.3           & 82.4 & 33.6 & $-$48.8 & 63.9 & 54.7 & $-$9.2 & 80.9 & 66.2 & $-$14.7 & 16.0 & 58.7 & +42.7 \\
OpenCode       & GPT-5.6 Terra     & 76.3 & 30.7 & $-$45.6 & 62.4 & 57.2 & $-$5.2 & 77.1 & 65.3 & $-$11.8 & 10.7 & 52.0 & +41.3 \\
OpenCode       & DeepSeek-v4.1-flash & 91.3 & 50.7 & $-$40.6 & 81.3 & 76.1 & $-$5.2 & 93.6 & 83.6 & $-$10.0 & 9.3 & 25.3 & +16.0 \\
OpenCode       & Qwen3.7-plus      & 77.4 & 15.8 & $-$61.6 & 54.4 & 50.8 & $-$3.6 & 82.9 & 58.9 & $-$24.0 & 21.3 & 69.3 & +48.0 \\
OpenCode       & GLM-5.3           & 86.7 & 41.8 & $-$44.9 & 65.6 & 64.0 & $-$1.6 & 83.1 & 69.5 & $-$13.6 & 13.3 & 50.7 & +37.4 \\
\bottomrule
\end{tabular}
\end{table*}
Table~\ref{tab:main} reports overall results where we draw several core insight:

\paragraph{Reuse leaves substantial room for improvement}
No configuration reliably builds on existing code. Even the strongest misses 24.3\% of repository targets and 13.6\% of its own earlier functions, although the requirement explicitly lists these functions and encourages reusing them; Qwen3.7-plus misses about half of the repository targets and a third of the self targets under both harnesses, leave a substantial room for improvement.

\paragraph{Reuse depends on exploration yet reading is not enough}
On the repository side, reuse tracks upstream exploration. Configurations that view at least 900 source lines per turn reach 47--61\% repository recall, while those viewing fewer than 700 lines reach only 35--42\%, and targets that were read are consistently more likely to be reused than targets that were not. On the self side, reading is not the bottleneck: every configuration reads at least 98.9\% of its own earlier targets, yet self reuse ranges only from 67.0\% to 86.4\%. The gap therefore lies in the decision to build on the code that was found, not in finding it.

\paragraph{Reuse aligns with both correctness and redundancy}
Downstream, reuse first aligns with correctness. Configurations that reuse more repository code generally also resolve more turns and pass more tests: DeepSeek-v4.1-flash leads on both reuse and correctness, while Qwen3.7-plus trails on both. Reuse also aligns with structure. Targets that are bypassed are far more likely than reused ones to have their logic re-implemented, so the two DeepSeek-v4.1-flash configurations, which reuse the most, have clearly the lowest $C_{\text{dup}}$ at 17.1\% and 21.1\%, while all others lie between 33.6\% and 46.7\%. Across configurations, then, reuse, correctness, and redundancy move together; the following sections show that within a configuration, as turns progress or the memory changes, reuse and redundancy shift while pass rates barely move.

\begin{table*}[t]
\centering
\small
\setlength{\tabcolsep}{4pt}
\caption{Effect of historical memory on OpenCode with Qwen3.7-plus (75 task chains, 375 turns per setting). Reuse, recall, correctness, and $C_{\text{dup}}$ are in \%; agent behavior and $C_{\text{dup}}$ are averaged over turns. Best reuse, recall, and correctness per column in \textbf{bold}.}
\label{tab:memory}
\begin{tabular}{@{}lcccccccccc@{}}
\toprule
& & & \multicolumn{4}{c}{Upstream} & \multicolumn{3}{c}{Downstream} \\
\cmidrule(lr){4-7}\cmidrule(lr){8-10}
& \multicolumn{2}{c}{Reuse} & \multicolumn{2}{c}{Recall} & \multicolumn{2}{c}{Agent Behavior} & \multicolumn{2}{c}{Correctness} & Redundancy \\
\cmidrule(lr){2-3}\cmidrule(lr){4-5}\cmidrule(lr){6-7}\cmidrule(lr){8-9}\cmidrule(lr){10-10}
Memory & repo & self & repo & self & Files & Lines & Resolve & Pass & $C_{\text{dup}}$ \\
\midrule
No memory        & \textbf{58.9} & 30.0 & \textbf{55.5} & 85.5 & 27.7 & 878 & \textbf{30.4} & \textbf{72.4} & 65.1 \\
Interface memory & 50.5 & \textbf{67.8} & 35.1 & \textbf{99.7} & 13.5 & 576 & \textbf{30.4} & 71.9 & 46.7 \\
Source memory    & 51.3 & 29.2 & 38.5 & 84.4 & 16.8 & 567 & 27.7 & 70.2 & 70.7 \\
\bottomrule
\end{tabular}
\end{table*}

\begin{table*}[t]
\centering
\small
\setlength{\tabcolsep}{4pt}
\caption{Effect of historical memory across turns on OpenCode with Qwen3.7-plus (\%). Self reuse is defined from turn~2; $\Delta$ is the change from the first to the last turn.}
\label{tab:memory-turns}
\begin{tabular}{@{}lcccccccccccc@{}}
\toprule
& \multicolumn{3}{c}{$\text{recall}_{\text{repo}}$} & \multicolumn{3}{c}{$\text{reuse}_{\text{repo}}$} & \multicolumn{3}{c}{$\text{reuse}_{\text{self}}$} & \multicolumn{3}{c}{$C_{\text{dup}}$} \\
\cmidrule(lr){2-4}\cmidrule(lr){5-7}\cmidrule(lr){8-10}\cmidrule(lr){11-13}
Memory & T1 & T5 & $\Delta$ & T1 & T5 & $\Delta$ & T2 & T5 & $\Delta$ & T1 & T5 & $\Delta$ \\
\midrule
No memory        & 77.0 & 46.5 & $-$30.5 & 56.4 & 61.4 & +5.0 & 34.2 & 32.1 & $-$2.1 & 22.7 & 89.3 & +66.6 \\
Interface memory & 77.4 & 15.8 & $-$61.6 & 54.4 & 50.8 & $-$3.6 & 82.9 & 58.9 & $-$24.0 & 21.3 & 69.3 & +48.0 \\
Source memory    & 77.8 & 20.6 & $-$57.2 & 53.4 & 50.1 & $-$3.3 & 29.6 & 29.1 & $-$0.5 & 17.3 & 94.7 & +77.4 \\
\bottomrule
\end{tabular}
\end{table*}

\subsection{Reuse Across Turns}

The aggregate results above average over all turns and hide how reuse evolves as a task chain grows. Since each turn adds the agent's own code to the workspace, later turns offer more opportunities to reuse and, equally, to bypass existing implementations. Table~\ref{tab:turns} therefore compares the first and last turn of each task chain.

\paragraph{Reuse declines over turns, especially for self targets}
As task chains progress, agents build less on existing code. Self reuse drops by 8--24 points from turn~2 to the last turn in every configuration, even though the requirement keeps listing the agent's earlier functions. The decline is steepest for Qwen3.7-plus, which loses 23--24 points, about twice the 10--13 points lost by DeepSeek-v4.1-flash.

\paragraph{Repository decline is driven by recall}
On the repository side, the decline in reuse follows the collapse of exploration. Agents read 76--91\% of repository targets at turn~1, but repository recall drops sharply once their own earlier code is available, reaching only 16--51\% by the last turn. Reuse falls far less than recall, so later-turn repository reuse is increasingly made without reading the target, plausibly from call sites seen in the agent's own earlier code. Self targets show the opposite pattern: recall stays saturated in every turn, so the decline in self reuse cannot be attributed to exploration.

\paragraph{Reuse decline leads to structural redundancy}
The decline in reuse has a structural consequence. When agents stop building on existing code, they re-implement it: targets that are bypassed are far more likely than reused ones to have their logic rewritten in the submission. Accordingly, $C_{\text{dup}}$ grows steadily over turns in every configuration, from 9--21\% of task chains at turn~1 to 51--69\% by the last turn; only DeepSeek-v4.1-flash, which retains the most self reuse, stays at 25--33\%. Because each turn inherits the workspace left by earlier turns, these re-implementations persist, leaving parallel versions of the same logic scattered across modules. None of this is visible in pass rates.

\subsection{Effect of Historical Memory}
\label{sec:memory}

In multi-turn development, what an agent is told about its earlier work may shape whether it builds on that work~\citep{packer2024memgptllmsoperatingsystems,li2025memosmemoryosai,zhang2024surveymemorymechanismlarge}. We therefore vary the form of historical memory in the requirement while keeping everything else fixed, using OpenCode with Qwen3.7-plus. Under all three settings, the agent's own earlier implementations remain in the workspace; only the requirement differs from turn~2 onward:
\begin{itemize}
    \item \textbf{No memory}: the requirement contains no information about earlier turns.
    \item \textbf{Interface memory} (our default): the requirement lists the module, name, signature, return value, and behavior of each function implemented in earlier turns, and encourages reusing them.
    \item \textbf{Source memory}: the requirement includes the complete source code of every earlier submission, verbatim.
\end{itemize}

Table~\ref{tab:memory} reports overall results under each memory setting, and Table~\ref{tab:memory-turns} compares their first and last turns.

\paragraph{A concise interface helps self reuse; full source code does not}
To get agents to build on their own earlier code, a short description of what already exists works, while handing them the complete source does not. Listing the interfaces of earlier functions raises self reuse from 30.0\% with no memory to 67.8\%, more than doubling it. Supplying the full source of earlier submissions instead leaves self reuse at 29.2\%, no better than giving no memory at all, and it stays flat at about 29\% in every turn. More information about earlier work is thus not what agents need; what helps is a compact map of which functions exist and where to find them. Interface memory mitigates rather than removes the problem, however: self reuse still declines over turns, from 82.9\% to 58.9\%. Note also that interface memory explicitly encourages reuse while source memory does not, so the two factors cannot be fully separated.

\paragraph{Memory substitutes for exploration}
Without memory, agents explore the workspace more broadly, opening 27.7 files per turn, about twice the 13.5 opened under interface memory, which raises repository recall from 35.1\% to 55.5\% and repository reuse from 50.5\% to 58.9\%. With interface memory, they rely on the note rather than exploring, and repository recall collapses twice as fast over turns, losing 61.6 points against 30.5 without memory. Memory thus shifts reuse between sources: it strengthens reuse of the agent's own code but partly at the expense of repository code that the note does not cover.

\paragraph{Memory changes redundancy but not correctness}
Correctness is nearly identical across the three settings, with resolved rates of 27.7--30.4\% and passed rates of 70.2--72.4\%, yet their reuse behavior differs sharply. Redundancy is lowest under interface memory, where $C_{\text{dup}}$ is 46.7\%, and highest under source memory, where it reaches 70.7\%: with the full source of earlier turns in context, agents tend to rewrite its logic rather than import it, and by the last turn 94.7\% of task chains contain a re-implemented target. As in the main results, a pass-rate-centered evaluation would treat these settings as equivalent, even though they leave very different workspaces behind.

\section{Conclusion}
We asked a question existing evaluation cannot answer---\emph{do coding agents reuse existing code or reinvent the wheel?} We built RepoReuse to make it measurable: a multi-turn benchmark on real repositories in which requirements are revealed turn by turn, the workspace accumulates across turns, and reuse of both pre-existing repository modules and the agent's own historical code is audited per turn. The benchmark is produced by a fully automated pipeline and scales readily to new repositories. The answer it gives is sobering: agents progressively stop exploring the repository they work in; they reuse their own history less even when it is fully in view, and handing them the complete source does not help; and the targets they bypass accumulate as cross-turn duplication that pass rates never register. Agents can solve every task and still degrade the codebase and pass rates cannot tell the difference. We hope RepoReuse provides a sustainable instrument for auditing, and ultimately improving, how agents build on existing code.

\newpage
\clearpage
\section*{Limitations}
RepoReuse currently covers five mature, pure-Python libraries and 75 five-turn task chains, evaluated with two harnesses and four backbone models; whether the observed deficiencies generalize to other languages, domains, and longer development horizons remains open. Widening the repository pool and the model--harness grid is work in progress, and the fully automated construction pipeline makes both straightforward to extend.


\bibliography{custom}

\appendix

\clearpage
\newpage
\section*{Appendix}
\label{sec:appendix}

\section{Experimental Setup Details}
\label{app:setup}

\subsection{Evaluation Protocol}
\label{app:protocol}

\paragraph{Turn Execution}
Each task chain runs in its own workspace, and its turns are executed in order. Before turn~$t$, the workspace contains the original repository together with every file the agent submitted in turns $1, \dots, t-1$, each at its required module path; reference solutions never enter the workspace. At each turn, the requirement is rendered according to the memory setting (Appendix~\ref{app:memory}) and handed to a fresh agent session with no dialogue history, trajectory, or test feedback from earlier turns. When the agent stops, whether by finishing, exhausting its budget, or failing, we read back whatever file exists at the required path and score it. A turn without a file at that path fails all of its tests. The hidden tests are copied into the workspace only after the agent stops, run from a temporary directory with a 600-second timeout, and removed afterwards. A failed turn does not end the chain: the next turn starts from whatever the agent actually left behind.

\paragraph{Workspace Preparation}
Before the first turn, we remove information that would let the agent bypass exploration. Git history is deleted, since \texttt{git log} would expose later changes. All test files are deleted file by file, including \texttt{test\_*.py}, \texttt{conftest.py}, and cached test listings, because repository tests are where the construction pipeline found its evidence and often spell out the conventions a task relies on. Test runners and helpers that the package imports at runtime are kept, and we verify that the package still imports after this step.

\paragraph{Environment}
Each repository is pinned to its benchmark commit and installed in editable mode in a dedicated conda environment, following the installation recipe of SWE-bench~\citep{jimenez2024swebenchlanguagemodelsresolve}. The same environment is used by all configurations, so that differences between runs come only from the agent.

\subsection{Harness and Model Configuration}
\label{app:harness}

\begin{table}[h]
\centering
\small
\setlength{\tabcolsep}{4pt}
\caption{Per-turn budgets and settings of the two harnesses.}
\label{tab:harness}
\begin{tabular}{@{}lcc@{}}
\toprule
& mini-SWE-agent & OpenCode \\
\midrule
Tools & bash only & native tools \\
Step limit & 120 commands & none \\
Cost limit & \$5 & none \\
Per-command timeout & 120\,s & harness default \\
Wall-time limit & 2,400\,s & 2,400\,s \\
Temperature & 0 & provider default \\
\bottomrule
\end{tabular}
\end{table}

\paragraph{mini-SWE-agent}
The agent interacts with the repository only through bash commands, each executed in a fresh subshell. Its prompt is adapted from the harness's default template with our delivery instructions (Appendix~\ref{app:memory}). We make one addition to the default observation template: each observation reports how many commands have been used, since without it agents tended to spend the whole budget exploring and submit nothing. The agent ends a turn with an explicit submit command. Table~\ref{tab:harness} lists its budgets; a turn is also stopped after three consecutive malformed responses.

\paragraph{OpenCode}
We run OpenCode in its headless mode with external plugins disabled and every tool call auto-approved. Each run uses an isolated configuration with a single model provider, so that no user-level settings, plugins, or agents affect the run. OpenCode has no built-in step or cost limit, so each turn is bounded only by the 2,400-second wall time. All models are accessed through the same OpenAI-compatible gateway under both harnesses.

\paragraph{Failures}
Agent errors, timeouts, and exceeded budgets are not retried. As described above, the turn is scored on whatever file the agent left at the required path, so such failures count against correctness rather than being excluded.

\subsection{Requirement and Memory Formats}
\label{app:memory}

\paragraph{Requirement Structure}
Every requirement follows the same structure: a behavioral description of the task, the exact path and importable module name of the file to create, and for each required function its signature, a description of its behavior, and a precise contract for its return value. From turn~2 onward, a historical note follows, whose form depends on the memory setting. Repository reuse targets are never named in the requirement.

\paragraph{Delivery Instructions}
Both harnesses wrap the requirement in the same delivery instructions, shown in Appendix~\ref{Prompts for RepoReuse Evaluation} for OpenCode; mini-SWE-agent uses the same wording within its own action protocol. These instructions are identical across all memory settings, so a generic encouragement to reuse existing code is present even when the historical note is removed or replaced.

\paragraph{Memory Settings}
The three memory settings of Section~\ref{sec:memory} differ only in the historical note. \emph{No memory} removes the note entirely. \emph{Interface memory}, our default, lists for every earlier turn its module and, for each function, its signature, return contract, and behavior, and encourages reusing them. \emph{Source memory} removes the interface note and instead embeds the exact source file submitted in each earlier turn, delimited by markers and labeled with its path. The box in Appendix~\ref{Prompts for RepoReuse Evaluation} shows the interface-memory note for turn~2 of a seaborn task, whose first turn implemented three color-mapping functions.

\section{Prompt Templates}
\label{Prompt Templates}
\subsection{Prompts for RepoReuse Pipeline}

\begin{promptbox}[breakable]{Prompt for Extracting Evidence Packages}
You are assembling an *evidence pack*: a small set of repository symbols that a programming task will be built on top of. You are walking a call graph, one hop at a time, like a developer clicking through cross-references.

Repository: {repo_name}
{round_ctx}
## Symbols collected so far
{collected}

## Current position
{current_block}

## Menu -- you may expand ONE of these, or stop
{menu}

## Your goal
Collect {lo}-{hi} symbols that could plausibly be combined into ONE coherent programming task (a small script with a few functions). Prefer symbols that genuinely belong together in a real workflow over symbols that are merely adjacent in the graph.

Reply with STRICT JSON only:
{{"action": "expand" | "stop",
  "choice": <menu number, required iff action=expand>,
  "reason": "<one sentence: why this symbol belongs with the ones collected>",
  "task_sketch": "<if stop: one sentence describing the task these symbols support>"}}
\end{promptbox}

\begin{promptbox}[breakable]{Prompt for Generating Tasks}
You are authoring a programming task for a benchmark that evaluates whether coding agents explore and reuse repository internals. Inputs: an evidence pack (repository symbols to build on), a suggested direction, the package layout, and previous-round artifacts.

Write a task a real developer would plausibly need, whose correct implementation GENUINELY requires reading the evidence symbols' source -- i.e., it must exploit internal conventions invisible in signatures (zero representation, coefficient ordering, implicit domain transforms). Any task solvable from docstrings alone is rejected.

Output six blocks:
1. module_name: snake_case; reads as if it always belonged to the package.
2. goal: 2-4 behavioural sentences, as a maintainer would request; never reference earlier rounds -- discovering what the repo already provides is what is being measured.
3. functions: {nfun} functions (name / signature / returns / behaviour); house-style names, no shared prefix; 5+ lines of real logic each; `returns` pins container, element type, and ordering; >=2 functions return plainly repr-comparable values.
4. solution: imports and calls the evidence symbols non-trivially; function names match block 3 verbatim; no module-level helpers.
5. calls: 8-14 single-line calls with predicted reprs, including edge cases; we execute each call and assert the real return value.
6. read_targets  locations a developer MUST read: the evidence symbols plus callees encoding conventions (even if never imported); each `why` names the convention, not the task.
\end{promptbox}
\subsection{Prompts for RepoReuse Evaluation}
\label{Prompts for RepoReuse Evaluation}

\begin{promptbox}{Delivery Instructions}
You are working in the repository checkout at {workdir} (your working directory). The package is installed in editable mode, so `python -c "import ..."` picks up any file you write there. The repository's own test suite is not available; write and run your own checks if you need to.

# Requirement

{requirement}

# What you must deliver

A new Python module at exactly `{workdir}/{submit_to}` (importable as `{module}`), defining exactly the functions the requirement lists, with those names and signatures. The requirement may mention modules that already exist in the repository -- find and reuse them where it makes sense; do not modify existing files.

Write the file, check that it imports and behaves as required, and fix what fails. A round with no file at that path scores zero however much you learned.\end{promptbox}

\begin{promptbox}{Interface Memory (turn 2 of a seaborn task)}
## Note

The repository already contains modules you implemented in earlier rounds. You are encouraged to reuse their public interfaces:

### Round 1 -- `seaborn._core.color_mapping`

- `color_lookup`
  - Signature: `def color_lookup(data, values=None, order=None)`
  - Returns: list[tuple[Any, tuple[float, ...]]] -- one entry per resolved category in seaborn order; NumPy scalar category labels are converted to equivalent Python scalars, and each color tuple contains exactly three RGB or four RGBA floats
  - Behaviour: Resolve the categorical levels and return the standardized color assigned to each level.
- `map_color_values`
  - Signature: `def map_color_values(data, values=None, order=None)`
  ...
- `describe_palette`
  - Signature: `def describe_palette(colors)`
  ...
\end{promptbox}

\begin{promptbox}{Source Memory (turn 2 of the same task)}
## Note

The following exact source files were submitted by this same harness in earlier rounds of this task. Byte lengths delimit source payloads.

### Round 1 -- `seaborn._core.color_mapping`

- Submit path: `seaborn/_core/color_mapping.py`
- UTF-8 bytes: <length>
- SHA-256: `<digest>`

<<<CODEFLOW_L5_SELF_SOURCE bytes=<length>>>>
[complete source of the round-1 submission]
<<<END_CODEFLOW_L5_SELF_SOURCE>>>
\end{promptbox}

\section{Case Study}
\begin{figure*}[t]
    \centering    \includegraphics[width=1.0\linewidth]{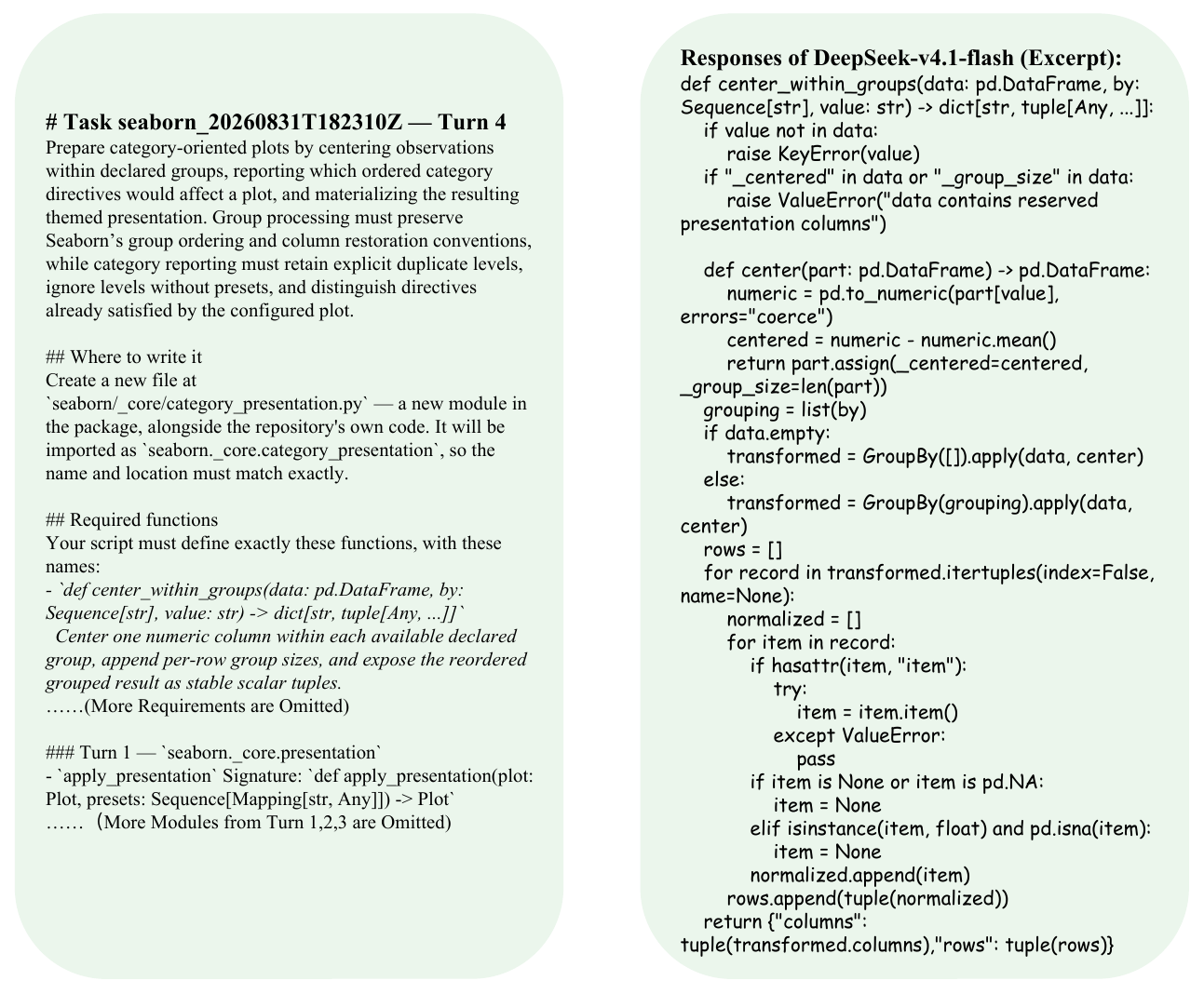}
    \caption{Case study, turn~4. \emph{Left:} the turn-4 requirement. \emph{Right:} the agent's \texttt{center\_within\_groups}, centering a numeric column within seaborn's categorical groups.}
    \label{Turn4}
\end{figure*}

\begin{figure*}[t]
    \centering    \includegraphics[width=1.0\linewidth]{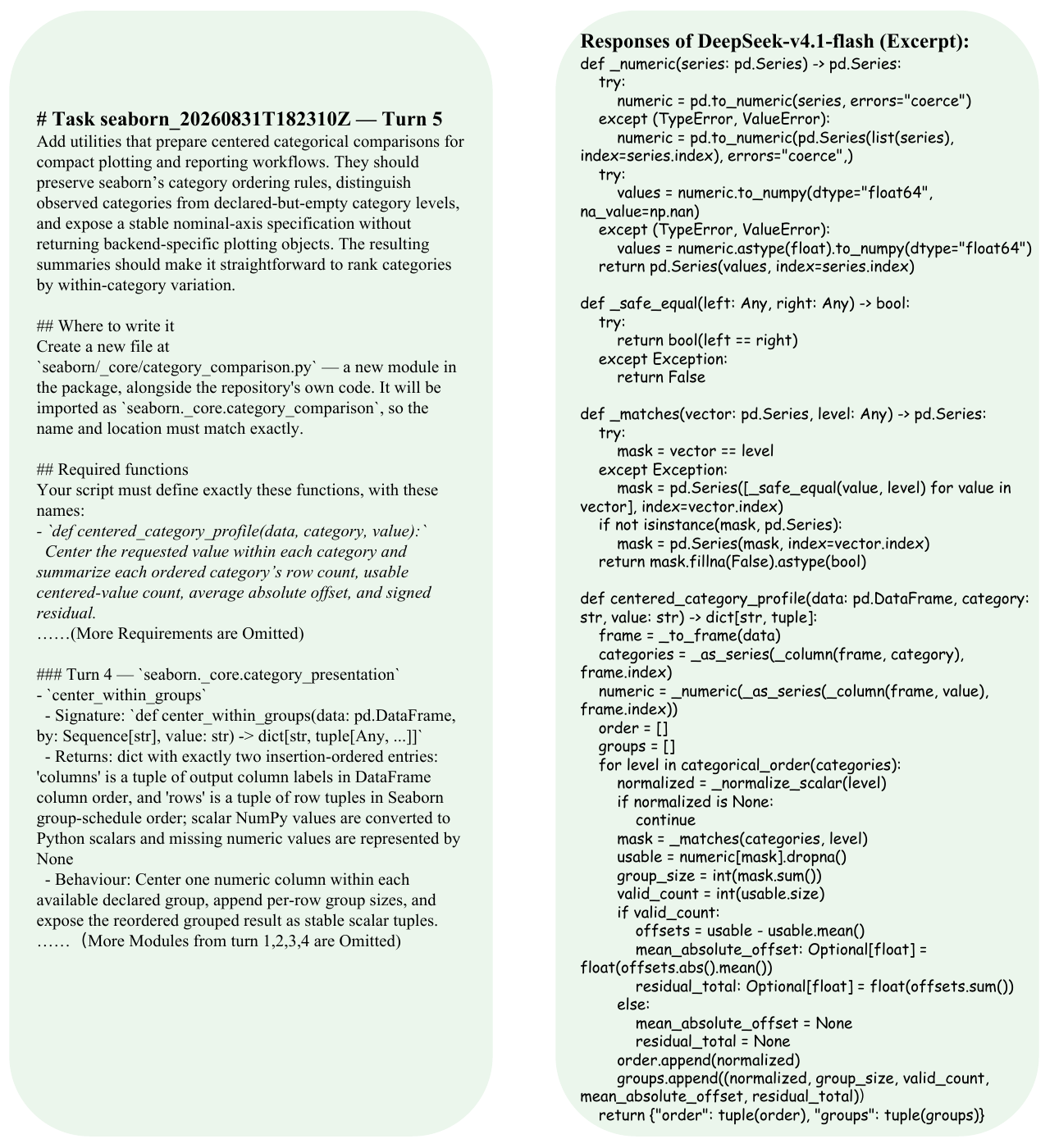}
    \caption{Case study, turn~5. \emph{Left:} the turn-5 requirement, which explicitly lists the turn-4 interface. \emph{Right:} the agent reads that code (recall $=1.0$) yet rewrites the same coercion, grouping, and centering logic from scratch instead of importing it---passing 10/10 tests while duplicating its own earlier implementation.}
    \label{Turn5}
\end{figure*}

Figures~\ref{Turn4} and~\ref{Turn5} show two consecutive turns of one task chain (mini-SWE-agent with DeepSeek-v4.1-flash). At turn~4, the agent implements \texttt{center\_within\_groups}, which centers a numeric column within seaborn's categorical groups, and passes all tests. At turn~5, the requirement asks for per-category summaries of centered values and explicitly lists the turn-4 interface. The agent reads the turn-4 source in full (recall $=1.0$), yet never imports it. Instead, it rewrites the dtype coercion, group masking, categorical ordering, and mean-centering logic from scratch in 262 lines, where the ground truth solution reaches the same behavior in 68 lines by calling the turn-4 function and aggregating its output. The submission passes 10/10 tests: nothing in the pass rate registers that the workspace now hosts two parallel implementations of the same centering semantics. 

\end{document}